\documentclass[preprint]{aastex}     

\usepackage{amsmath}    
\usepackage{lineno}     
\usepackage{xspace}     

\shorttitle{Comparing Cosmic Ray Energy Spectra}
\shortauthors{BenZvi et al.}

\begin{document}

\title{A Bayesian Approach to Comparing Cosmic Ray Energy Spectra}

\author{S.Y. BenZvi\altaffilmark{1}, B.M. Connolly\altaffilmark{2}, C.G. Pfendner\altaffilmark{1}, 
and S. Westerhoff\altaffilmark{1}}

\altaffiltext{1}{Department of Physics, University of Wisconsin-Madison, Madison, WI 53706}
\altaffiltext{2}{University of Pennsylvania, Philadelphia, PA 19104}

\begin{abstract}

A common problem in ultra-high energy cosmic ray physics is the comparison of
energy spectra.  The question is whether the spectra from two experiments or
two regions of the sky agree within their statistical and systematic
uncertainties.  We develop a method to directly compare energy spectra for
ultra-high energy cosmic rays from two different regions of the sky in the same
experiment without reliance on agreement with a theoretical model of the energy
spectra.  The consistency between the two spectra is expressed in terms of a
Bayes factor, defined here as the ratio of the likelihood of the two-parent
source hypothesis to the likelihood of the one-parent source hypothesis.
Unlike other methods, for example $\chi^{2}$ tests, the Bayes factor allows 
for the calculation of the posterior odds ratio and correctly accounts for 
non-Gaussian uncertainties.  The latter is particularly important at the highest
energies, where the number of events is very small.

\end{abstract}


\section{Introduction}

A century after Victor Hess's discovery of cosmic rays, it is still unclear
where and how these particles are accelerated.  Some of them reach energies
above $10^{20}$\,eV, well above the capabilities of man-made
accelerators.  (See~\cite{Beatty:2009zz} for a recent review.) Clues about
the origin of these ultra-high energy cosmic rays comes from the study of the
cosmic ray energy spectrum.  While basically a simple power law over the entire
range of measured energies from GeV to above EeV, the spectrum shows some
important features that might hold the key to discovering and understanding the
sources.  The most relevant feature at the ultra-high energy end of the
spectrum is the flux suppression above around $6\times 10^{19}$\,eV caused by
the interaction of cosmic ray primaries with the photons of the 2.7\,K
microwave background, the so-called Greisen-Zatsepin-Kuz\'min (GZK)
suppression~\citep{Greisen:1966jv,Zatsepin:1966jv}, predicted already in 1966
shortly after the discovery of the microwave background.  Recently,
a flux suppression at the highest energies, consistent with the GZK suppression,
has been observed in data recorded by the High Resolution Fly's Eye experiment
in Utah~\citep{Abbasi:2007sv} and the Pierre Auger Observatory in
Argentina~\citep{Abraham:2008ru}.  If this suppression is indeed the long-sought
GZK suppression and not an intrinsic feature of the sources,
we now know that most of the highest energy cosmic rays are
produced at large distances.  Those observed on Earth with energies above
$6\times 10^{19}$\,eV must originate from sources closer than 80 to 100\,Mpc,
or from within what is referred to as the ``GZK sphere.''

The existence of a suppression at ultra-high energies is not all that can be
learned from the energy spectrum.  The exact shape of the spectrum in the GZK
suppression region can provide information on the actual distribution of the
sources.  Furthermore, the energy spectrum is sensitive to a variety of
factors, including production and transport mechanisms, and cosmic ray mass
composition.  Because of the sensitivity of the spectrum to these effects, it
is useful to examine the spectrum in multiple ways.  Experiments like the
Pierre Auger Observatory and the Telescope Array experiment now collect data at 
an unprecedented rate, so several studies that were not possible years ago when 
the total number of detected events at the highest energies was
little more than a handful, are now possible for the first time.

One possible study that may give some insight into the origin of cosmic rays is
a comparison of the energy spectrum in different regions of the sky.  The
spectrum in a region that contains one or more strong cosmic ray sources
can potentially deviate from the all-sky spectrum.  If the source is
closer than 80\,Mpc, for example, its flux is not expected to show a GZK 
suppression.  Recently, the region around the Active Galactic Nuclei (AGN) 
Centaurus A has been identified as a possible region of an enhanced cosmic 
ray flux in Auger data~\citep{Abraham:2007si,Abreu:2010zzj}.
Since Cen\,A is nearby (4\,Mpc), the energy spectrum in the Cen\,A region 
could differ from the all-sky cosmic ray energy spectrum. 

More generally, increased statistics from the current generation of instruments
will eventually allow a detailed comparison of the shape of the cosmic ray flux 
as a function of the sky position, thus creating a ``skymap'' of spectral 
parameters, for example of the spectral index.  Such a study
might reveal sky regions where cosmic ray accelerators are located.  For this
study to be as general as possible, it should not be limited to comparing power
law indices, as the spectrum in certain parts of the sky might not be well
described by a power law or even a broken power law.  An ideal method would compare
the shape of the spectrum in a certain region of the sky to the all-sky flux
without any prejudice as to the functional form of the spectrum. 

An additional complication is the fact that measurements of the energy spectrum
are often plagued by 20\%-30\% systematic uncertainties in energy measurement
and low statistics at the highest energies.  The measurements at the highest
energy values are often determined by only a few events.  A rigorous
statistical analysis must therefore be applied to the spectra to compare them
and extract any sort of meaning. 

In this paper, we develop a statistical method to compare cosmic ray energy
spectra.  The method uses a Bayes factor formulation where the likelihood of
the hypothesis that the two energy spectra stem from one source (the
``one-parent'' hypothesis) is compared to the likelihood of the hypothesis that
the spectra stem from from different sources (the ``two-parent'' hypothesis).

There are several advantages to a Bayesian approach to model selection.  Most
importantly, it allows for the calculation of the posterior odds ratio in favor 
of the two-parent hypothesis over the one-parent hypothesis, which is the 
relevant model selection parameter.  Unlike a $\chi^2$-test, it takes into 
account the alternative hypothesis, and it automatically penalizes over-fitting 
of the data with complex models.  In contrast to
a $\chi^2$- or $F$-test, it allows for non-Gaussian uncertainties in the data,
a feature that is important in the comparison of cosmic ray energy spectra, as
the number of events at the highest energies is very small.
In addition, the Bayesian formalism allows for the marginalization of nuisance
parameters and systematic uncertainties.  Marginalization provides a convenient
way to quantify our ignorance of nuisance parameters with the judicious choice
of prior probability distributions. 

We develop two different techniques for comparing the spectra.  The first
method compares the absolute flux of the spectra.  This method depends on
knowledge of the relative exposure of the two data sets.  The second method is
similar except that we remove the dependence on the known relative exposure
and compare the spectra using no absolute scale; instead, we marginalize the
relative weight (the scaling factor) of the spectra in the one-parent case.
This lack of dependence on the relative exposure allows one to compare the
shape of the spectra without comparing the absolute flux.  This is useful in
cases where the relative exposure between data sets is not known with
sufficient accuracy, or when the absolute flux is not considered relevant in
the comparison.

The paper is organized as follows.  In Section 2, we develop the two methods to
compare the energy spectra.  In Section 3, we use simulated data to test
the methods and evaluate their sensitivity.  In Section 4, we compare the 
Bayes factor method to a $\chi^2$-test.  The paper is summarized in Section 5.

\section{Method}

Let $\vec{\mathcal{F}}_{1}=\{\mathcal{F}_{1,i}\}$ and
$\vec{\mathcal{F}}_{2}=\{\mathcal{F}_{2,i}\}$ be the (binned) observed fluxes
that are to be compared.  Then the Bayes factor $B_{21}$ is the
likelihood ratio that the measurements arise from two parent distributions
versus a single parent distribution,
\begin{equation}
B_{21}=\frac{P(\vec{\mathcal{F}}_{1},\vec{\mathcal{F}}_{2}|H_{2})}
{P(\vec{\mathcal{F}}_{1},\vec{\mathcal{F}}_{2}|H_{1})}~~,
\label{eq:bayes1}
\end{equation}
where $H_{1}$ and $H_{2}$ indicate the one- and two-parent hypothesis, 
respectively.  The Bayes factor is equal to the posterior odds ratio
\begin{equation}
\frac{P(H_2|\vec{\mathcal{F}}_{1},\vec{\mathcal{F}}_{2})}
     {P(H_1|\vec{\mathcal{F}}_{1},\vec{\mathcal{F}}_{2})} = 
\frac{P(\vec{\mathcal{F}}_{1},\vec{\mathcal{F}}_{2}|H_{2})P(H_2)}
{P(\vec{\mathcal{F}}_{1},\vec{\mathcal{F}}_{2}|H_{1})P(H_1)} =
     B_{21}\frac{P(H_2)}{P(H_1)}~~,
\label{eq:postodds}
\end{equation}
commonly used in Bayesian model selection when $P(H_1)=P(H_2)$, {\it i.e.}, 
when the prior probabilities of the hypotheses in question are equivalent
\citep{KassRaftery:1995,Goodman:1999b}.  In other words, $B_{21}$ is a quantity
which derives solely from the data.  It describes how
the data will cause an experimenter to favor one or another hypothesis after
conducting an experiment, independent of prior beliefs or prejudices regarding
the two hypotheses. 

Because it is a ratio, $B_{21}$ can take on any value between $0$ and $\infty$.
To make sense of its value, it is convenient to note the connection between the
Bayes factor and the posterior probability.  For example, if we do not favor
either model before taking data ($P(H_1)=P(H_2)$), we can use Bayes' Theorem to
express the posterior probability of the null (one-parent) hypothesis purely
in terms of the Bayes factor:
\begin{equation}
P(H_1|\vec{\mathcal{F}}_{1},\vec{\mathcal{F}}_{2}) =
\frac{1}{1 + B_{21}}
\label{eq:postbf}
\end{equation}
A Bayes factor $B_{21}=10^{-2}$ indicates that the posterior probability of the
one-parent hypothesis given the data is $99\%$; and $B_{21}=10^2$ indicates
that the posterior probability is $0.99\%$.  Hence, it is conventional to
interpret $B_{21}>10^2$ as strong or decisive evidence against the null
hypothesis, and $B_{21}<10^{-2}$ as decisive evidence in favor of the null
hypothesis~\citep{Jeffreys:1939gh}.  Of course, it is possible to adjust these
decision thresholds according to one's needs.  If it is preferable to use the
``5-sigma'' convention of overwhelming evidence, the necessary limits on the
Bayes factor can be computed using Eq.~\eqref{eq:postbf}, assuming there are
no prior prejudices toward $H_1$ or $H_2$.

In this calculation, we compare the observed fluxes $\vec{\mathcal{F}}_{1}$ and
$\vec{\mathcal{F}}_{2}$ with the expected values $\vec{f}_{1}=\{f_{1,i}\}$ and
$\vec{f}_{2}=\{f_{2,i}\}$ given a particular hypothesis.  It is convenient to
express the expectation in terms of the total expected counts $\{\eta_{i}\}$ 
and a set of weights $\{w_{i}\}$ such that
\begin{equation}
f_{1,i}=w_{i} \eta_{i},~~~~~f_{2,i}=(1-w_{i}) \eta_{i}~.
\end{equation}
The weights have values between 0 and 1 and the counts can take on any positive 
value.  Since the expected fluxes $\vec{f}_{1}$ and $\vec{f}_{2}$ are unknown, we
marginalize these parameters in the Bayes factor, so Eq.\,\ref{eq:bayes1} becomes
\begin{eqnarray}
B_{21} & = & \frac{\int\int
P(\vec{\mathcal{F}}_{1},\vec{\mathcal{F}}_{2}|\vec{f}_{1},\vec{f}_{2},H_2)
P(\vec{f}_{1},\vec{f}_{2}|H_{2}) d\vec{f}_{1} d\vec{f}_{2}}
{\int\int
P(\vec{\mathcal{F}}_{1},\vec{\mathcal{F}}_{2}|\vec{f}_{1},\vec{f}_{2},H_1)
P(\vec{f}_{1},\vec{f}_{2}|H_{1}) d\vec{f}_{1} d\vec{f}_{2}}\\
            & = & \frac{\int\int
            P(\vec{\mathcal{F}}_{1},\vec{\mathcal{F}}_{2}|\vec{w'},\vec{\eta},H_2)
P(\vec{w'},\vec{\eta}|H_{2}) d\vec{w'} d\vec{\eta}}
{\int\int P(\vec{\mathcal{F}}_{1},\vec{\mathcal{F}}_{2}|\vec{w},\vec{\eta},H_1)
P(\vec{w},\vec{\eta}|H_{1}) d\vec{w} d\vec{\eta}}~.
\label{eq:bayes2}
\end{eqnarray}
Method A and B now differ in the treatment of the weights $w_i$ for the one-
and two-parent hypotheses.  These prior model restrictions on the weights can
be introduced via the probabilities $P(\vec{f}_{1},\vec{f}_{2}|H_{1})$ and
$P(\vec{f}_{1},\vec{f}_{2}|H_{2})$.

\subsection{Method A: Comparing Absolute Flux}

In method A, in the one-parent hypothesis, the weights are simply the
(known) relative exposure for the two data sets:
\begin{equation}
w_{i}=\frac{(\text{Exposure}_1)_{i}}{(\text{Exposure}_1)_{i}+(\text{Exposure}_2)_{i}}~.
\end{equation}
In the denominator of Eq.\,\ref{eq:bayes2}, the marginalization over the
weights $w_{i}$ therefore collapses since we equate the weights with the
relative experimental exposures.  In the two-parent hypothesis, every possible
relative exposure is allowed since the absolute flux in the two regions could
be different. Therefore, each of the weights are allowed to take on any value
between 0 and 1.

We treat the error in the flux as Poissonian, so the probability
$P(\vec{\mathcal{F}}_{1},\vec{\mathcal{F}}_{2}|w,\eta)_{i}$ of observing counts
$\{\mathcal{F}_{1,i}\}$ and $\{\mathcal{F}_{1,i}\}$ given expected counts
$\vec{f}_{1}=\{w_{i}\eta_{i}\}$ and $\vec{f}_{2}=\{(1-w_{i})\eta_{i}\}$ becomes
\begin{equation}
P(\mathcal{F}_{1},\mathcal{F}_{2}|w,\eta)_{i} = 
\frac{(w_{i}\eta_{i})^{\mathcal{F}_{1,i}} e^{-w_{i}\eta_{i}}}{\mathcal{F}_{1,i}!}
\frac{((1-w_{i})\eta_{i})^{\mathcal{F}_{2,i}} e^{-(1-w_{i})\eta_{i}}}{\mathcal{F}_{2,i}!}~~,
\end{equation}
thus the Bayes factor is
\begin{equation}
B_{21}=\frac{\prod_{i=1}^{N}{\int^{1}_0{dw_{i}} \int^{\infty}_{0} {d\eta_i} 
\frac{(w_{i}\eta_{i})^{\mathcal{F}_{1,i}} e^{-w_{i}\eta_{i}}}{\mathcal{F}_{1,i}!}
\frac{((1-w_{i})\eta_{i})^{\mathcal{F}_{2,i}} e^{-(1-w_{i})\eta_{i}}}
{\mathcal{F}_{2,i}!}}}
{\prod_{i=1}^{N}{\int^{\infty}_{0}{d\eta_{i}} \frac{(w_{i}\eta_{i})^{\mathcal{F}_{1,i}}
e^{-w_{i}\eta_{i}}}{\mathcal{F}_{1,i}!}
\frac{((1-w_{i})\eta_{i})^{\mathcal{F}_{2,i}} e^{-(1-w_{i})\eta_{i}}}{\mathcal{F}_{2,i}!}}}~,
\end{equation}
where $N$ is the number of bins.  Note that a flat prior 
$P(\eta_{i}|H)=1/(\eta_\text{max}-\eta_\text{min})$ 
for $\eta_{i}$ is actually improper in the limit $\eta_\text{min}=0$ and 
$\eta_\text{max}\rightarrow\infty$.  The problem can be circumvented by 
explicitely using $\eta_\text{min}$ and $\eta_\text{max}$ and letting 
them go to zero and infinity only after integration~\citep{jaynes2003ptl}.  
In our example, the $\eta_{i}$-dependence actually cancels out.

Rearranging the Bayes factor, one gets
\begin{equation}
B_{21}=\frac{\prod_{i=1}^{N}{\int^{1}_{0}{dw_{i}}
w_{i}^{\mathcal{F}_{1,i}}(1-w_{i})^{\mathcal{F}_{2,i}}
\int^{\infty}_{0}{d\eta_{i}} e^{-\eta_{i}}\eta_{i}^{\mathcal{F}_{1,i}+\mathcal{F}_{2,i}}}}
{\prod_{i=1}^{N}{w_{i}^{\mathcal{F}_{1,i}}(1-w_{i})^{\mathcal{F}_{2,i}}
\int^{\infty}_{0}{d\eta_{i}} e^{-\eta_{i}}\eta_{i}^{\mathcal{F}_{1,i}+\mathcal{F}_{2,i}}}}~.
\end{equation}
Since the $\eta_i$ terms cancel, this reduces to
\begin{equation}
B_{21}=\frac{\prod_{i=1}^{N}{\int^{1}_{0}{d w_{i}} w_{i}^{\mathcal{F}_{1,i}}
(1-w_{i})^{\mathcal{F}_{2,i}}}} {\prod_{i=1}^{N}{w_{i}^{\mathcal{F}_{1,i}}
(1-w_{i})^{\mathcal{F}_{2,i}}}}~.
\end{equation}
Using the identity
\begin{equation}
\frac{\Gamma(a)~\Gamma(b)}{\Gamma(a+b)} = \int_{0}^{1} t^{a-1}(1-t)^{b-1}dt~,
\label{eq:gamma}
\end{equation}
the Bayes factor can be written as
\begin{equation} \label{eq:GammaA}
B_{21}=\frac{\prod_{i=1}^{N}{\frac{\Gamma(\mathcal{F}_{1,i}+1)~\Gamma(\mathcal{F}_{2,i}+1)}
{\Gamma(\mathcal{F}_{1,i}+\mathcal{F}_{2,i}+2)}}}
{\prod_{i=1}^{N}{w_{i}^{\mathcal{F}_{1,i}}(1-w_{i})^{\mathcal{F}_{2,i}}}}~.
\end{equation}

In our case, the $\mathcal{F}$ terms are all positive integers, so the gamma
functions reduce to factorials and we are left with the following form:
\begin{equation}
B_{21}=\frac{\prod_{i=1}^{N}\frac{\mathcal{F}_{1,i}!~~\mathcal{F}_{2,i}!}
{(\mathcal{F}_{1,i}+\mathcal{F}_{2,i}+1)!}}
{\prod_{i=1}^{N} w_{i}^{\mathcal{F}_{1,i}}(1-w_{i})^{\mathcal{F}_{2,i}}}~.
\end{equation}
For the purpose of calculation, it is more convenient to deal with the logarithm of
the Bayes factor in Eq.\,\ref{eq:GammaA},
\begin{equation} 
\begin{split} 
\ln B_{21}=\sum_{i=1}^N \Bigl[\ln(\Gamma(\mathcal{F}_{1,i}+1))+\ln(\Gamma(\mathcal{F}_{2,i}+1))-\ln(\Gamma(\mathcal{F}_{1,i}+\mathcal{F}_{2,i}+2))\\
-\mathcal{F}_{1,i}\ln(w_{i})-\mathcal{F}_{2,i}\ln(1-w_{i})\Bigr]~~.
\end{split}
\end{equation}

\subsection{Method B: Comparing Shape of Spectrum Only}

Next we want to compare the shape of the spectra without making any assumptions
on the relative exposure.  This is relevant in cases
where we do not want the comparison to depend on an
accurate knowledge of the exposure.  The two-parent case remains the same as in
method A, since we already allow every possible relative exposure.  However,
the one-parent hypothesis needs to be modified.  We now allow the weights
$w_{i}$ to float, but not from bin to bin as in method A.  Rather, we want the 
weight to act as a normalization factor to allow the spectra to scale together
over all bins at once.  The weights $w_{i}$ are therefore not bin-dependent
and can be described by a single weight $w=w_{i}$ which is allowed to float
between 0 and 1.

The Bayes factor therefore now becomes
\begin{equation}
B_{21}=\frac{\prod_{i=1}^{N}{\int^{1}_{0}{dw_{i}} \int^{\infty}_{0}{d\eta_i} 
\frac {(w_{i}\eta_{i})^{\mathcal{F}_{1,i}} e^{-w_{i}\eta_{i}}}{\mathcal{F}_{1,i}!} 
\frac{((1-w_{i})\eta_{i})^{\mathcal{F}_{2,i}} e^{-(1-w_{i})\eta_{i}}} {\mathcal{F}_{2,i}!}}}
{\int^{1}_{0}{dw}\prod_{i=1}^{N}{\int^{\infty}_{0}{d\eta_{i}} 
\frac{(w\eta_i)^{\mathcal{F}_{1,i}} e^{-w\eta_{i}}}{\mathcal{F}_{1,i}!}}
\frac{((1-w)\eta_{i})^{\mathcal{F}_{2,i}} e^{-(1-w)\eta_{i}}}{\mathcal{F}_{2,i}!}}~.
\end{equation}
Again, the $\eta_{i}$ terms cancel and $B_{21}$ reduces to
\begin{equation}
B_{21}=\frac{\prod_{i=1}^{N}{\int^{1}_{0}{dw_{i}} w_{i}^{\mathcal{F}_{1,i}}
(1-w_{i})^{\mathcal{}F_{2,i}}}}
{\int^{1}_{0}{dw} w^{\sum_{i=1}^{N}\mathcal{F}_{1,i}} 
(1-w)^{\sum_{j=1}^{N}\mathcal{F}_{2,j}}}~.
\end{equation}
Since the sums $\sum_{i=1}^{N}\mathcal{F}_{1,i}$ and
$\sum_{i=1}^{N}\mathcal{F}_{2,i}$ are simply the total number of events $N_{1}$
and $N_{2}$ in spectrum 1 and 2, this becomes 
\begin{equation} \label{eq:GammaB}
B_{21}=\left(\prod_{i=1}^{N}{\frac{\Gamma(\mathcal{F}_{1,i}+1)~\Gamma(\mathcal{F}_{2,i}+1)}
{\Gamma(\mathcal{F}_{1,i}+\mathcal{F}_{2,i}+2)}}\right)
\frac{\Gamma(N_{1}+N_{2}+2)}{\Gamma(N_{1}+1)~\Gamma(N_{2}+1)}~,
\end{equation}
which simplifies to
\begin{equation}
B_{21}=\left(\prod_{i=1}^{N}{\frac{\mathcal{F}_{1,i}!~\mathcal{F}_{2,i}!}
{(\mathcal{F}_{1,i}+\mathcal{F}_{2,i}+1)!}}\right)
\frac{(N_{1}+N_{2}+1)!} {N_{1}!~N_{2}!}~.
\end{equation}
As before, we actually the logarithm of the Bayes factor in Eq.\,\ref{eq:GammaB}, 
\begin{equation} 
\begin{split}
\ln B_{21}=\sum_{i=1}^N{\Bigl[\ln(\Gamma(\mathcal{F}_{1,i}+1))+\ln(\Gamma(\mathcal{F}_{2,i}+1))-\ln(\Gamma(\mathcal{F}_{1,i}+\mathcal{F}_{2,i}+2))\Bigr]} \\
{+\ln(\Gamma(N_{1}+N_{2}+2))-\ln(\Gamma(N_{1}+1))-\ln(\Gamma(N_{2}+1))~~.}
\end{split}
\end{equation}

\section{Sensitivity} 

In this section, we evaluate the sensitivity of the
methods by appyling them to simulated spectra.  We start with a few simple
examples, comparing single power law spectra with different spectral indices,
and single and broken power laws.  These examples are meant to illustrate the
general behavior of the Bayes factor.  We will then study the sensitivity of
the methods for more realistic scenarios, for example for an analysis that
compares the energy spectrum in the region around a potential source to the
all-sky cosmic ray energy spectrum.  Several features of the spectra we compare
in this section will closely resemble the shape of the most recent published
energy spectrum of the Pierre Auger Observatory~\citep{Abraham:2010mj}.  To
summarize, the spectrum exhibits two main features, the ``ankle'' at
$\log(E_\text{ankle}/\mathrm{eV})=18.61\pm0.01$, and the onset of a flux
suppression at $\log(E_\text{br}/\mathrm{eV})=19.46\pm0.03$.  At the ankle, the
energy spectrum flattens from a spectral index of $\gamma_1=3.26\pm0.04$ to
$\gamma_2=2.59\pm0.02$.  At the suppression, the spectrum steepens again to
a spectral index $\gamma_3=4.3\pm0.2$.  The data is binned in 20 bins from
$\log(E/\mathrm{eV})=18.4$ to $\log(E/\mathrm{eV})=20.4$.
In this paper, we focus on the energy
spectrum above the ankle, which contains 14\,519 events recorded with the
surface detector array. 

As described in the previous section, a Bayes factor $B_{21}>1$ indicates that
the two-parent hypothesis is supported, but only larger Bayes factors
$B_{21}>10$ or $B_{21}>10^2$ provide substantial or decisive evidence against
the one-parent hypothesis.  Here, we will typically
require the Bayes factor to exceed $B_{21}>10^2$, considering the region
$10^{-2}<B_{21}<10^2$ as an ``undecided'' region, {\it i.e.}, a region where the
evidence is too weak to come to a conclusion for or against the two-parent
hypothesis.

For the following studies, we simulate power law spectra assuming Poissonian
errors on the number of events per energy bin.  As described in Section 2, we
calculate the Bayes factor based on the number of events per energy bin, $N_i$,
rather than the flux per energy bin.  The spectral indices for the number of
events $N$ versus energy and flux versus energy differ by 1, so a spectral
index of $\gamma=2.7$ for the flux (roughly the measured all-sky value)
corresponds to an index of 1.7 for the number of events.

\subsection{Comparing Two Single Power Law Spectra}

\begin{figure}
\epsscale{0.8}
\plotone{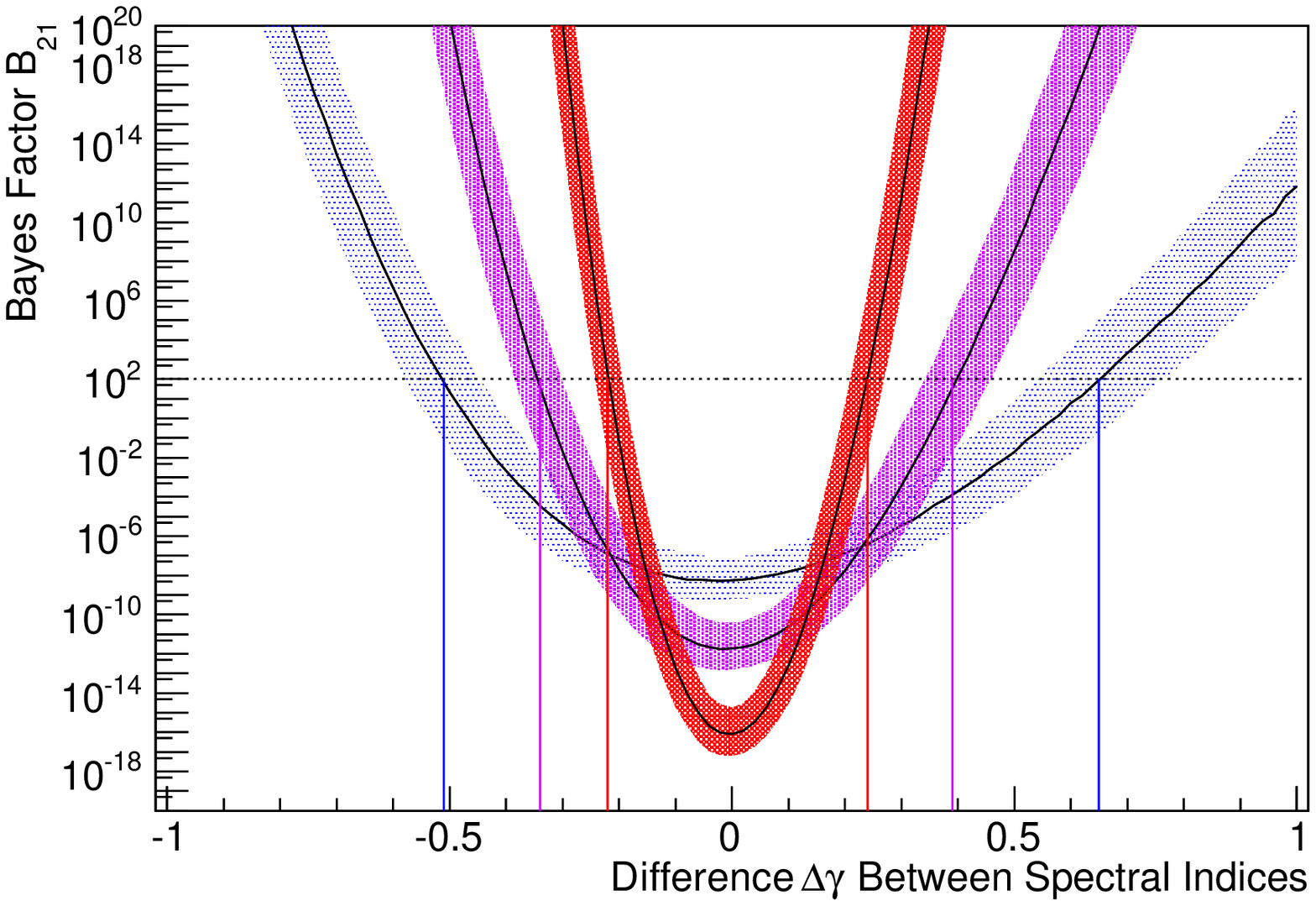}
\plotone{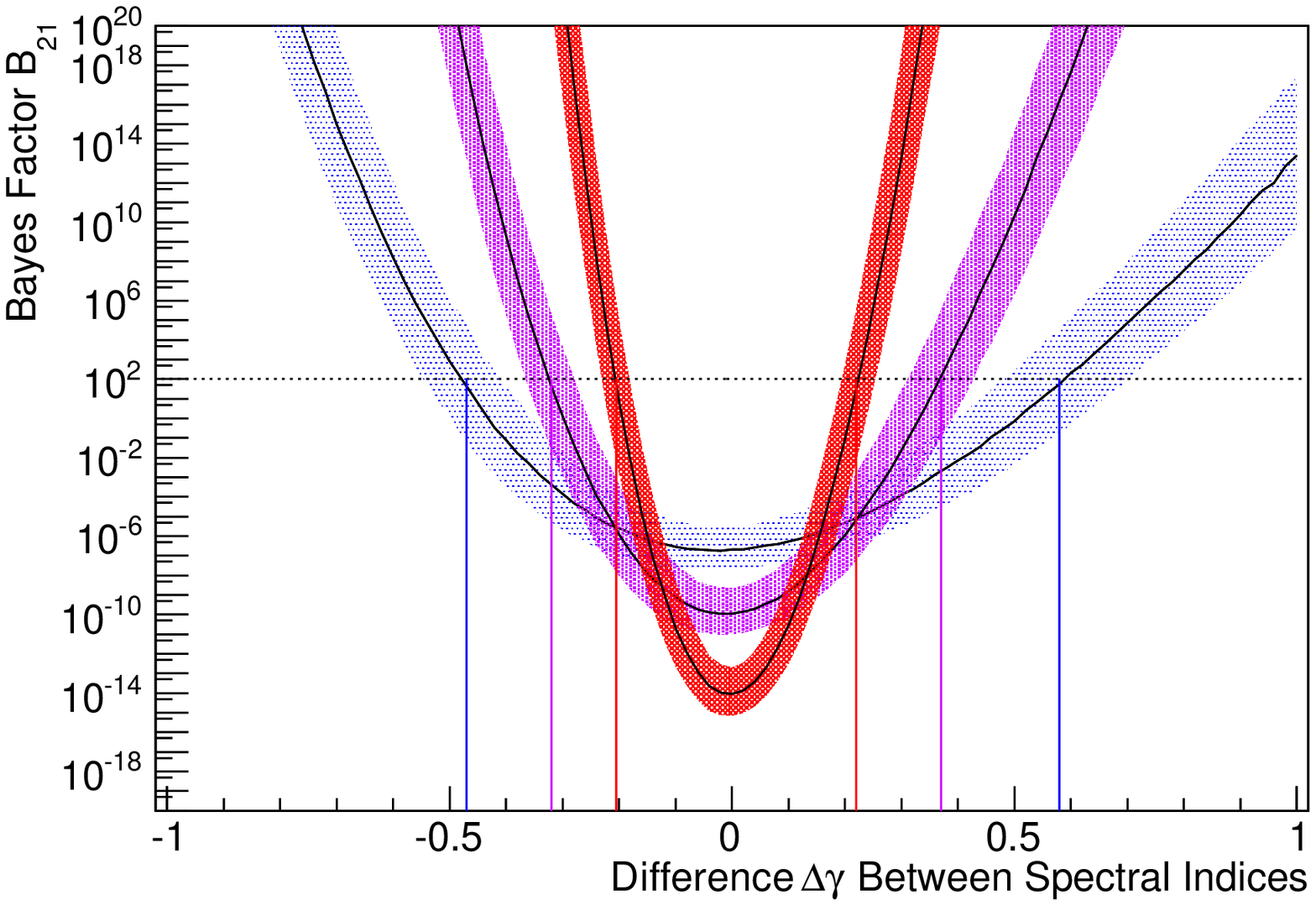}
\caption{
Bayes factor $B_{21}$ as a function of difference in spectral index,
$\Delta\gamma=\gamma_2-\gamma_1$, for two power law spectra, analysed with
method A {\it (top)} and method B {\it (bottom)}.  The spectral index of the
first set is $\gamma_1=2.7$.  The number of events in each set is 1000 (blue),
3000 (violet), and 10\,000 (red), and the shaded area represents the 68\,\% error
in each set.  The analysis is performed for a large number of random
implementations of the two data sets.  The solid line indicates the median and
the shaded area the 68\,\% percentile.  The horizontal line indicates a Bayes
factor $B_{21}=100$.
\label{fig:R_vs_dg}}
\end{figure}

We first compare two simulated power law spectra with spectral indices
$\gamma_1$ and $\gamma_2$, respectively.  The ability of any method to separate
two spectra with a difference $\Delta\gamma=\gamma_2-\gamma_1$ in spectral
indices will depend on the number of events in each data set.  To illustrate
the general behavior of the Bayes factor, we first compare two simulated data
sets of equal size $N$, but different spectral indices.  The spectral index of
the first data set is $\gamma_1=2.7$, and the spectral index of the second data
set is $\gamma_2$.  Fig.\,\ref{fig:R_vs_dg} shows the Bayes factor
$B_{21}$ as a function of the difference $\Delta\gamma=\gamma_2-\gamma_1$
for three different data set sizes and both methods.  For $\Delta\gamma=0$, the
two spectra are identical, and the Bayes factor takes on small values,
indicating strong support for the one-parent hypothesis.  As expected, the
support for the one-parent hypothesis is strongest for the largest data set
size.  For increasing and decreasing values of $\Delta\gamma$, the Bayes factor
quickly rises, and a Bayes factor of $B_{21}>100$, indicating
significant evidence that the data sets have different spectral indices, is
reached faster for the larger data sets.  The difference in spectral indices
that the methods can resolve decreases from about 0.5 for data sets with
$N=1000$ to 0.2 for data sets with $N=10\,000$.  To show the statistical 
error of the Bayes factor determination, here and in the following analyses, 
the calculation of the Bayes factor is performed for a large number of random
implementations of the two data sets, and the plot shows the median Bayes
factor and the band that contains 68\,\% of the random implementations.

\begin{figure}
\epsscale{0.8} 
\plotone{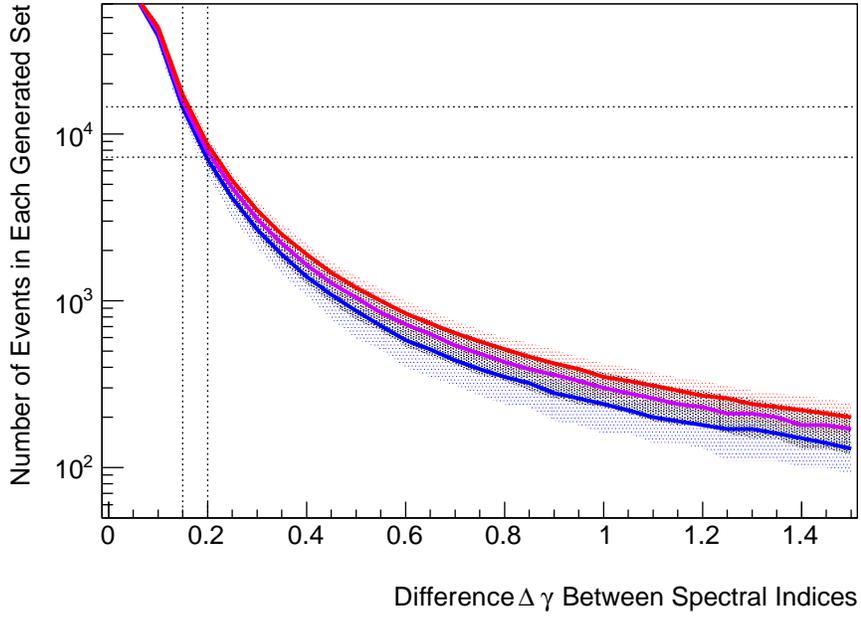}
\plotone{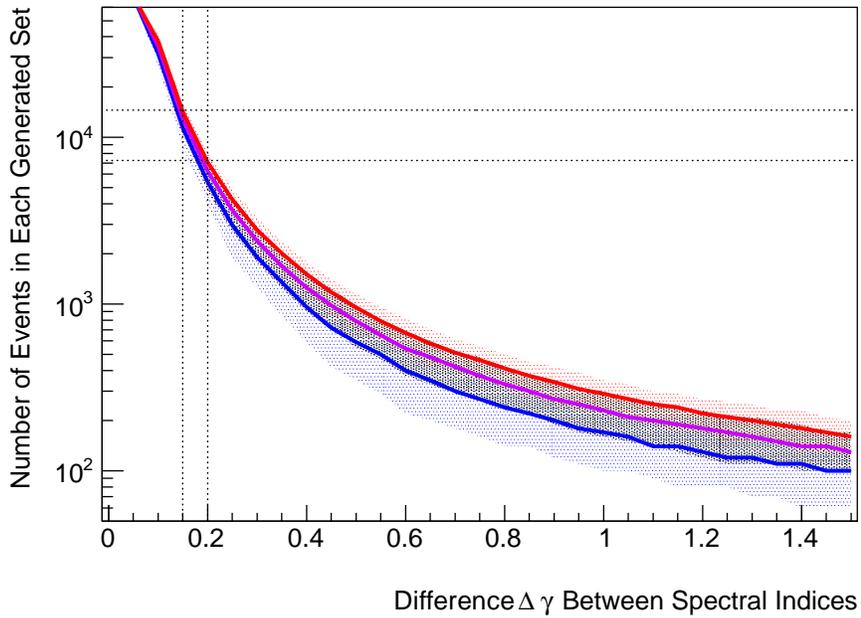}
\caption{
Number of events in each set required for two data sets representing single power
laws with different spectral indices $\gamma_1$ and $\gamma_2$ so the Bayes factor
$B_{21}$ reaches 100 (blue), 1000 (violet), and 10\,000 (red), as a
function of the difference $\Delta\gamma=\gamma_2-\gamma_1$ in spectral index.
The spectral index of the first data set is fixed at $\gamma_1=2.7$.  Results are
shown for method A {\it (top)} and method B {\it (bottom)}.  The analysis
is performed for a large number of random implementations of the two data sets.
The solid line indicates the median and the shaded area the 68\,\% percentile.
The horizontal lines represent the number of events in the Auger data set and half
that value.  The vertical lines indicate differences in the power law index of 0.15
and 0.2.
\label{fig:N_vs_dg}}
\end{figure}

An important question is how small the difference $\Delta\gamma$ can be
before the method can no longer differentiate between the two power spectra,
{\it i.e.} before the Bayes factor drops below some minimum value.
The smallest $\Delta\gamma$ that the method can resolve with $B_{21}$
above the desired minimum value is a measure of its sensitivity.  It depends 
on the size of the data sets, with larger data sets improving the sensitivity.
It also depends on the desired minimum Bayes factor, {\it i.e.} on 
the strength of the evidence against the one-parent hypothesis that the 
analyser requires.  Fig.\,\ref{fig:N_vs_dg} shows the number of
events necessary to reach Bayes factors $B_{21}=$100, 1000, and
10\,000 as a function of $\Delta\gamma$ for both methods.  As an example, for a
data set with 10,000 events in each set, the methods can resolve differences
in $\Delta\gamma$ of less between 0.2.  To reach a resolution of 0.15, around
15,000 events are necessary.  For this analysis, the 
sensitivities for methods A and B are roughly identical. 

\subsection{Comparing a Single Power Law Spectrum to a Broken Power Law Spectrum}
\label{subsec:broken}

Another simple example is the comparison of two spectra where one is a single
power law and the other is a broken power law.  This is an example with
important applications.  If we assume that the spectral index of the second
data set is identical to the index of the first data set for energies below
some break energy $E_\text{br}$ and different at energies above $E_\text{br}$,
this example describes a scenario where a GZK-type suppression is present in
one data set, but not in the other.  This could potentially be the case for the
comparison of the energy spectrum in the vicinity of a strong source of
ultra-high energy cosmic rays to the all-sky cosmic ray spectrum if the source
is within the GZK sphere and its flux is not subject to a suppression.

We start with a simple comparison using two data sets of the same size.  
The first data set is a single power law with spectral index $\gamma_1=2.59$.
The second data set is a broken power law with the same index $\gamma_1$ from
the lowest energy bin $\log(E/\mathrm{eV})=18.61$ to the break energy
$\log(E_\text{br}/\mathrm{eV})=19.46$, and a steeper index $\gamma_2=4.3$ above
$E_\text{br}$.  This shape corresponds to the spectrum measured by the Pierre
Auger Observatory~\citep{Abraham:2010mj}.

The spectra are produced in such a way that the total number of events below
$E_\text{br}$ is identical for the two data sets, so the data sets differ (in
the differential and integral number of events) only above $E_\text{br}$.  
Both data sets contain about 10000 events, but because it has more events at
higher energies, the data set following the single power law contains about 
200 events more.

\begin{figure}
\epsscale{1.0}
\plotone{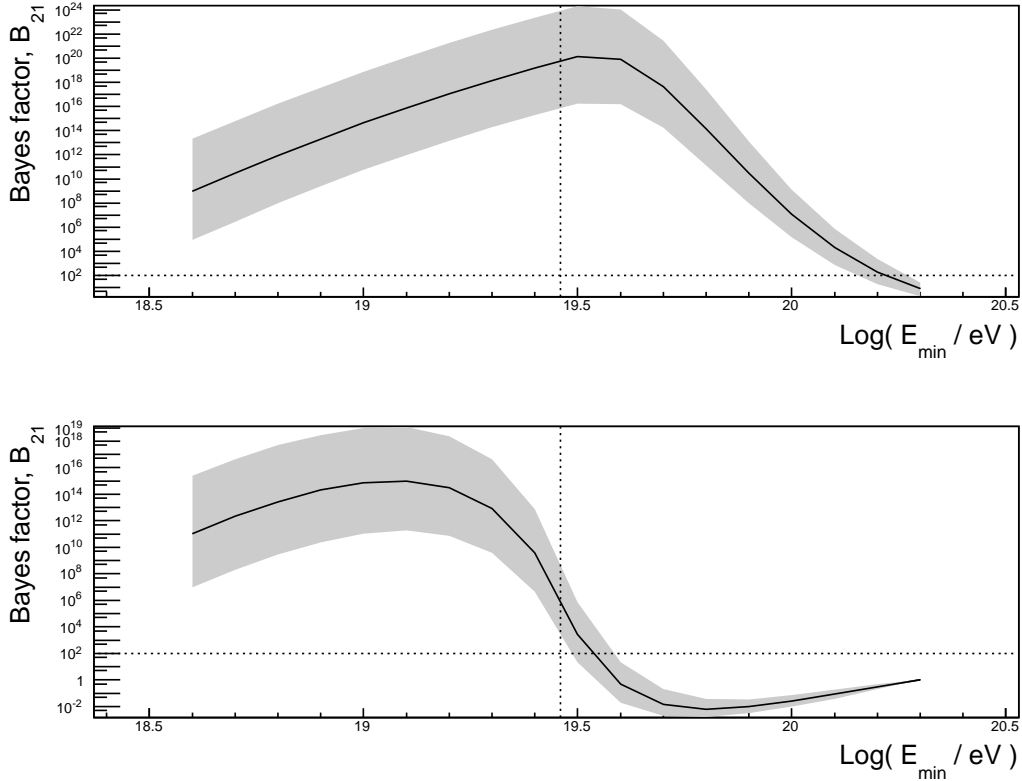}
\caption{
Bayes factor as a function of lower energy threshold $E_\text{min}$ for a comparison
between a single power law and broken power law for method A {\it (top)} and B
{\it (bottom)}.  The two spectra are identical within statistical errors for
energies below $\log(E_\text{br}/\mathrm{eV})=19.46$.  Above $E_\text{br}$, the
second data set steepens from $\gamma_1=2.59$ to $\gamma_2=4.3$.  Each sets contains
10000 events.  The analysis is performed for a large number of random
implementations of the two data sets.  The solid line indicates the median and
the shaded area the 68\,\% percentile.  The horizontal line indicates a Bayes
factor $B_{21}=100$. The vertical line indicates the position of the
breakpoint in the broken power law at $\log(E_\text{br}/\mathrm{eV})=19.46$.
\label{fig:single_broken}}
\end{figure}

The Bayes factor will depend on the energy range considered for the comparison.  
We consider the energy range from some lower energy threshold $E_\text{min}$ to 
the highest energies.  Fig.\,\ref{fig:single_broken} shows the Bayes factor as 
a function of the lower energy threshold $E_\text{min}$.  The Bayes factor 
increases with increasing $E_\text{min}$ and reaches a maximum.  For method A, 
which compares the spectra in shape and in absolute flux, the maximum Bayes 
factor occurs at about $\log(E_\text{min}/\mathrm{eV})=19.6$, slightly above
$\log(E_\text{br}/\mathrm{eV})=19.46$.  This behavior is expected, as the
spectra below the break agree and therefore do not contribute to the Bayes
factor.  For method B, which compares shape only, the Bayes factor reaches its
maximum at a lower energy, around $\log(E_\text{min}/\mathrm{eV})=19.2$, indicating
that more data is necessary for method B to reach the maximum of discrimination
power.  This is not surprising; since method B examines shape only, it
requires more low-energy bins to recognize a difference between the spectra.
Method A relies in part on the relative exposure, which like the spectral
index is different above $\log(E_\text{br}/\mathrm{eV})=19.46$.  After
the maximum is reached, the Bayes factor decreases with $E_\text{min}$ as the
data sets become smaller and can no longer be distinguished due to
low statistics.

\subsection{Prospects for Studies of the Spectrum as a Function of Sky Location}

In the study of the spectrum from a region around a strong source, the two data
sets to be compared will typically be unequal in size, reflecting the fact that
a small sky region around the source position is compared to the
rest of the sky.  A realistic test should account for the difference in
the sizes of the data sets.  We repeat the previous analysis comparing a single power
law to a broken power law, but now the total number of events $N_\text{tot}$
is distributed unequally: the single power law, representing the source region,
contains (as an example) 5\,\% of all events, whereas the broken power law,
representing the all-sky cosmic ray flux, contains 95\,\% of all events.

\begin{figure}
\epsscale{1.0}
\plotone{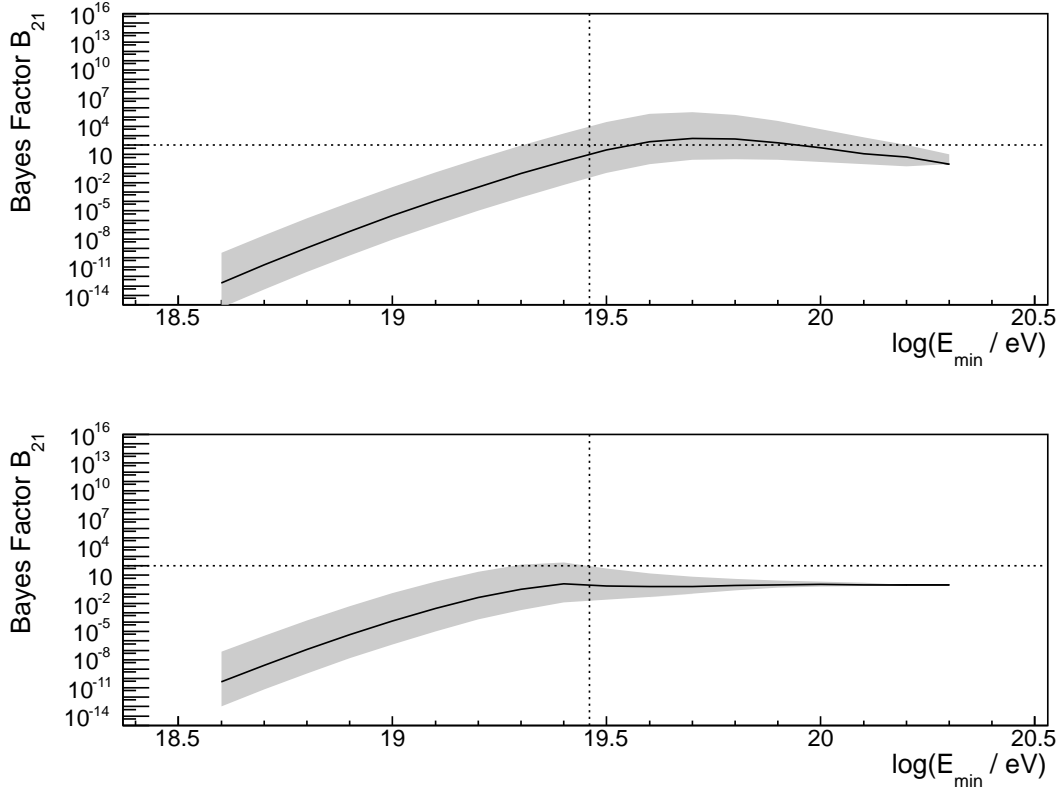}
\caption{
Bayes factor as a function of lower energy threshold $E_\text{min}$ for a comparison
between a single power law and broken power law for method A {\it (top)} and B
{\it (bottom)}.  The spectra have the same shape as in
Fig.\,\ref{fig:single_broken}, but here, the total number of events in both
sets is $N_\text{tot}=14519$, with the broken power law data set containing
$0.95\times N_\text{tot}$ events and the single power law data set containing
$0.05\times N_\text{tot}$ events.  The analysis is performed for a large number of
random implementations of the two data sets.  The solid lines indicate the
median and the shaded area the 68\,\% percentile.  The horizontal line
indicates a Bayes factor $B_{21}=100$. The vertical line indicates the
position of the breakpoint in the broken power law at
$\log(E_\text{br}/\mathrm{eV})=19.46$.
\label{fig:single_broken_5a}}
\end{figure}

\begin{figure}
\epsscale{1.0}
\plotone{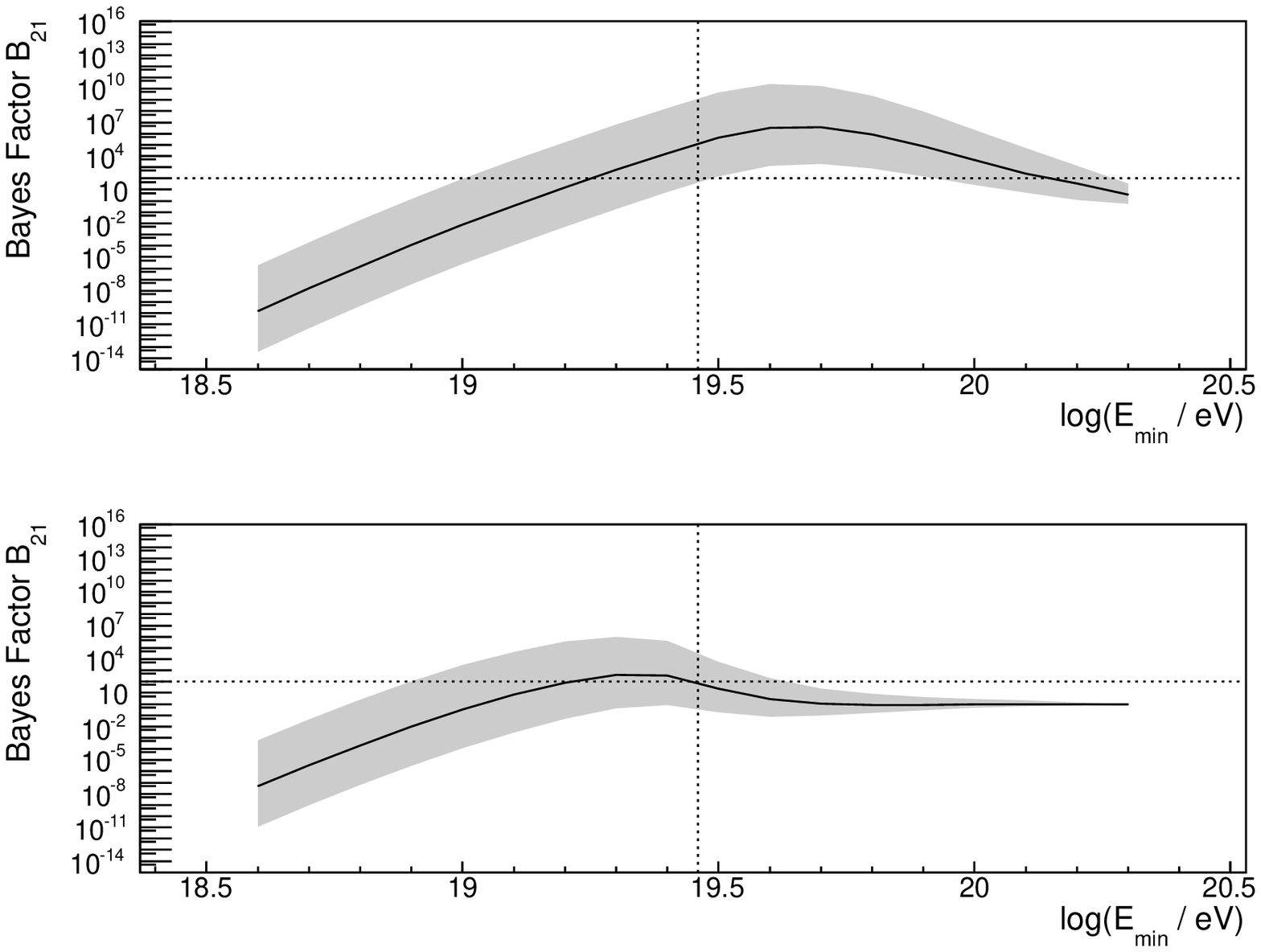}
\caption{Same as Fig.\,\ref{fig:single_broken_5a}, but for $N_\text{tot}=29038$.
\label{fig:single_broken_5b}}
\end{figure}

\begin{figure}
\epsscale{1.0}
\plotone{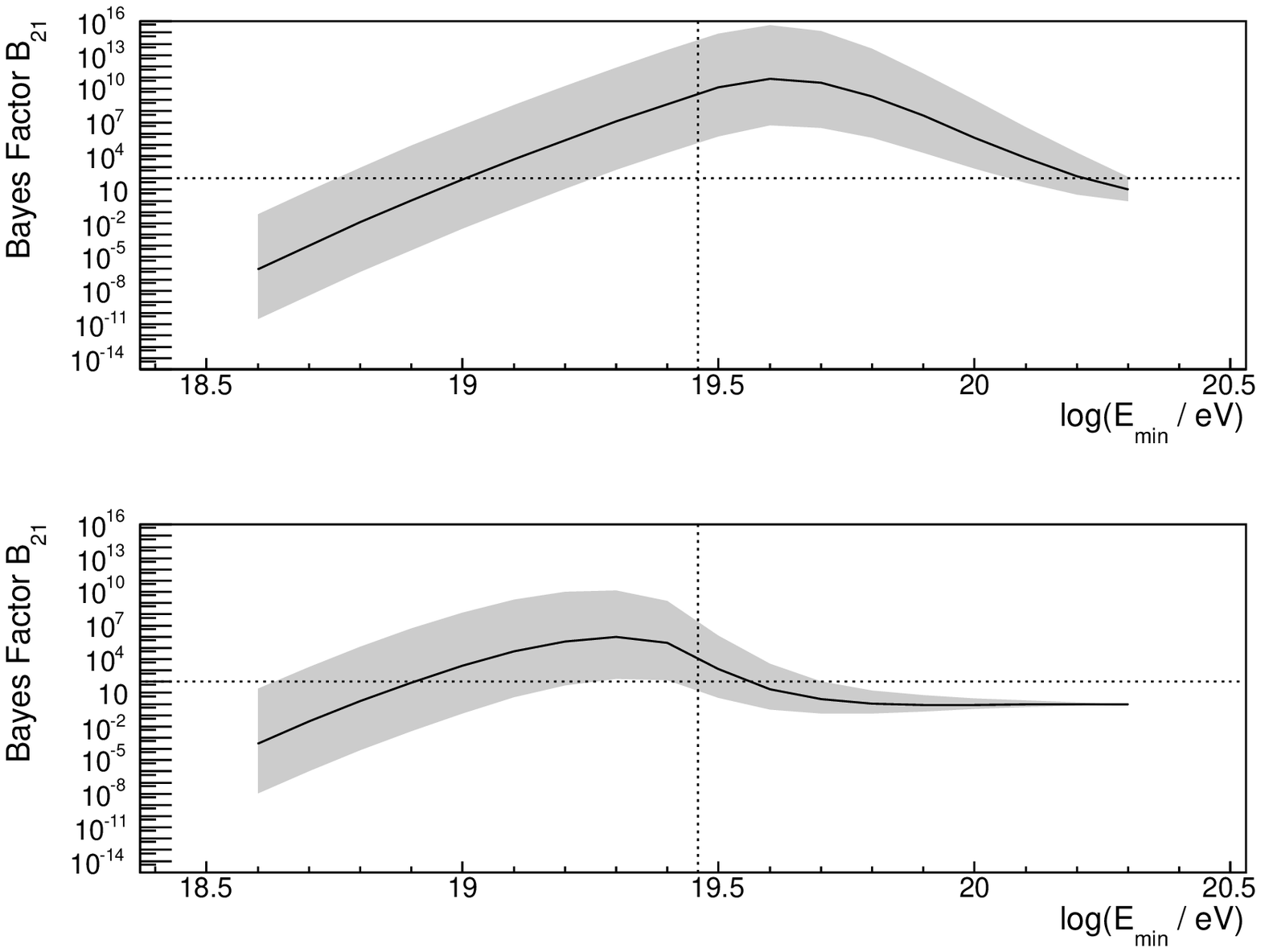}
\caption{Same as Fig.\,\ref{fig:single_broken_5a}, but for $N_\text{tot}=43557$. 
\label{fig:single_broken_5c}}
\end{figure}

The Bayes factor as a function of the lower energy threshold is shown in
Fig.\,\ref{fig:single_broken_5a}, with $N_\text{tot}=14519$, for method A 
(upper plot) and method B (lower plot).  We also show the results for a
data set with twice (Fig.\,\ref{fig:single_broken_5b}) and three times the
number of events (Fig.\,\ref{fig:single_broken_5c}) to illustrate the
improvement expected for larger data sets within reach of Auger during its
anticipated lifetime.

The analysis indicates that with the current data, only method A can 
discriminate between the two spectral shapes with a Bayes factor exceeding
100.  As in the example discussed in Section\,\ref{subsec:broken},
the Bayes factor reaches a peak value at
energies slightly higher than $E_\text{br}$ for method A, whereas method B
requires more data below the break energy.  The Bayes factor increases with
the size of the data set, and for a data set of twice the 
published size, method B starts to reach Bayes factors above 100 on average.
The Bayes factors increase to peak values of $10^{10}$ and $10^{5}$ for method
A and B, respectively, for data sets of three times the published
size.  The data recorded with the fully-operational Pierre Auger 
Observatory should reach this size within three to four years.

\begin{figure}
\epsscale{1.0}
\plotone{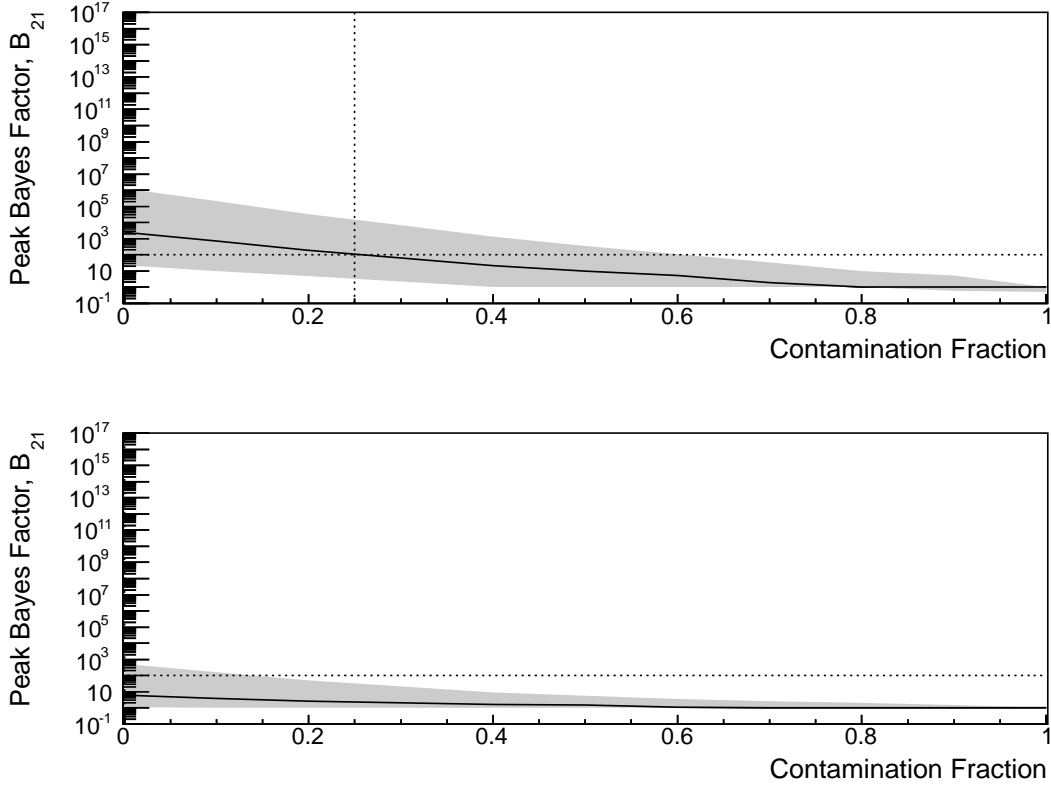}
\caption{
Peak Bayes factor as a function of the contamination fraction of a single power
law source by the all-sky background broken power law spectrum for method A
{\it (top)} and B {\it (bottom)}.  The spectra have the same shape as in
Fig.\,\ref{fig:single_broken}. The total number of events is $N_\text{tot}=14519$,
with the broken power law data set containing $0.95\times N_\text{tot}$ events and
the single power law data set containing $0.05\times N_\text{tot}$ events.  The
analysis is performed for a large number of random implementations of the two
data sets.  The solid lines indicate the median and the shaded area the 68\,\%
percentile.  The horizontal line indicates a Bayes factor $B_{21}=100$.
The vertical line indicates the contamination fraction for which the median of the 
peak Bayes factor starts to exceed 100.
\label{fig:single_broken_5a_reduced}}
\end{figure}

\begin{figure}
\epsscale{1.0}
\plotone{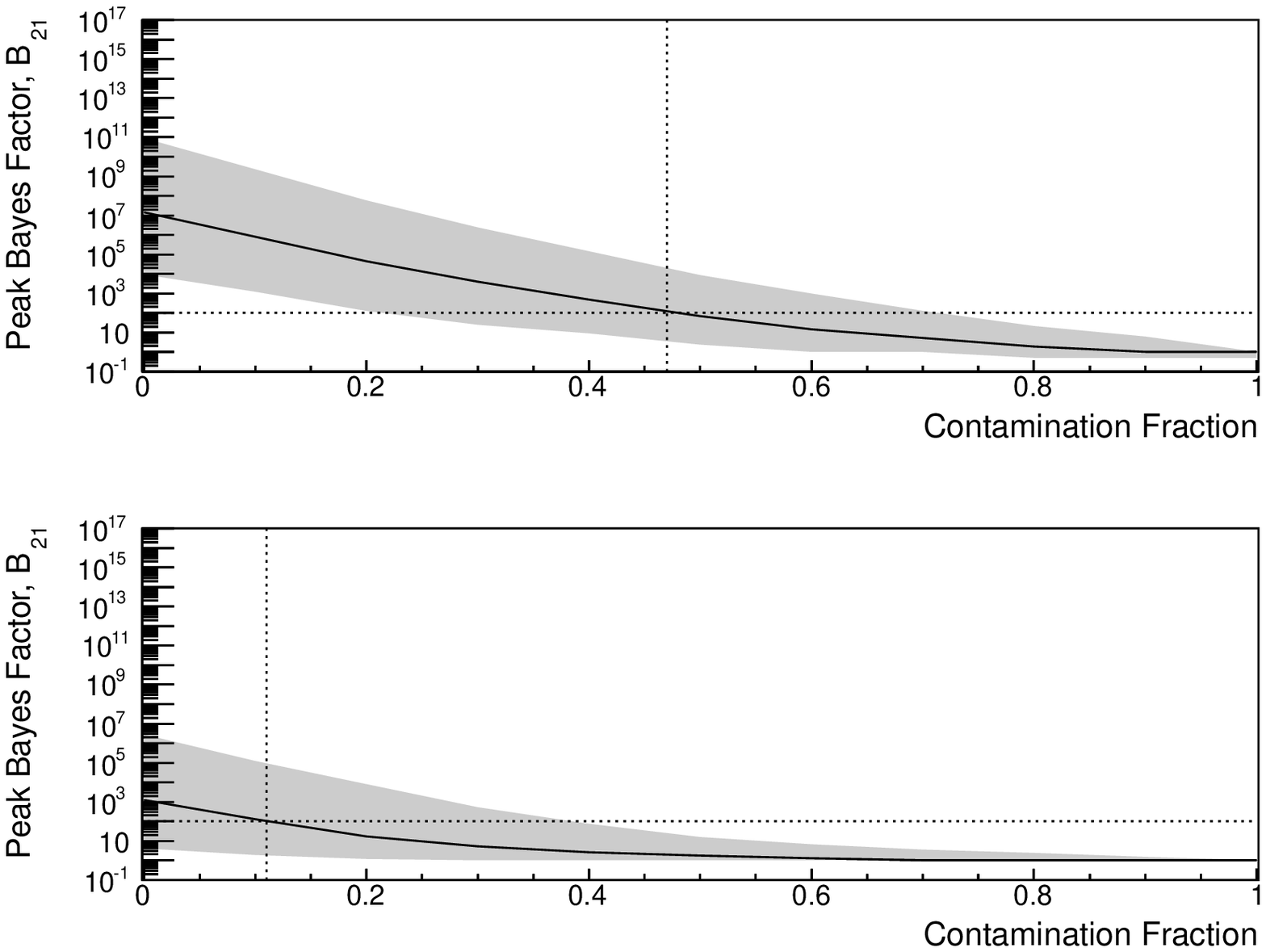}
\caption{Same as Fig.\,\ref{fig:single_broken_5a_reduced}, but for $N_\text{tot}=29038$.
\label{fig:single_broken_5b_reduced}}
\end{figure}

\begin{figure}
\epsscale{1.0}
\plotone{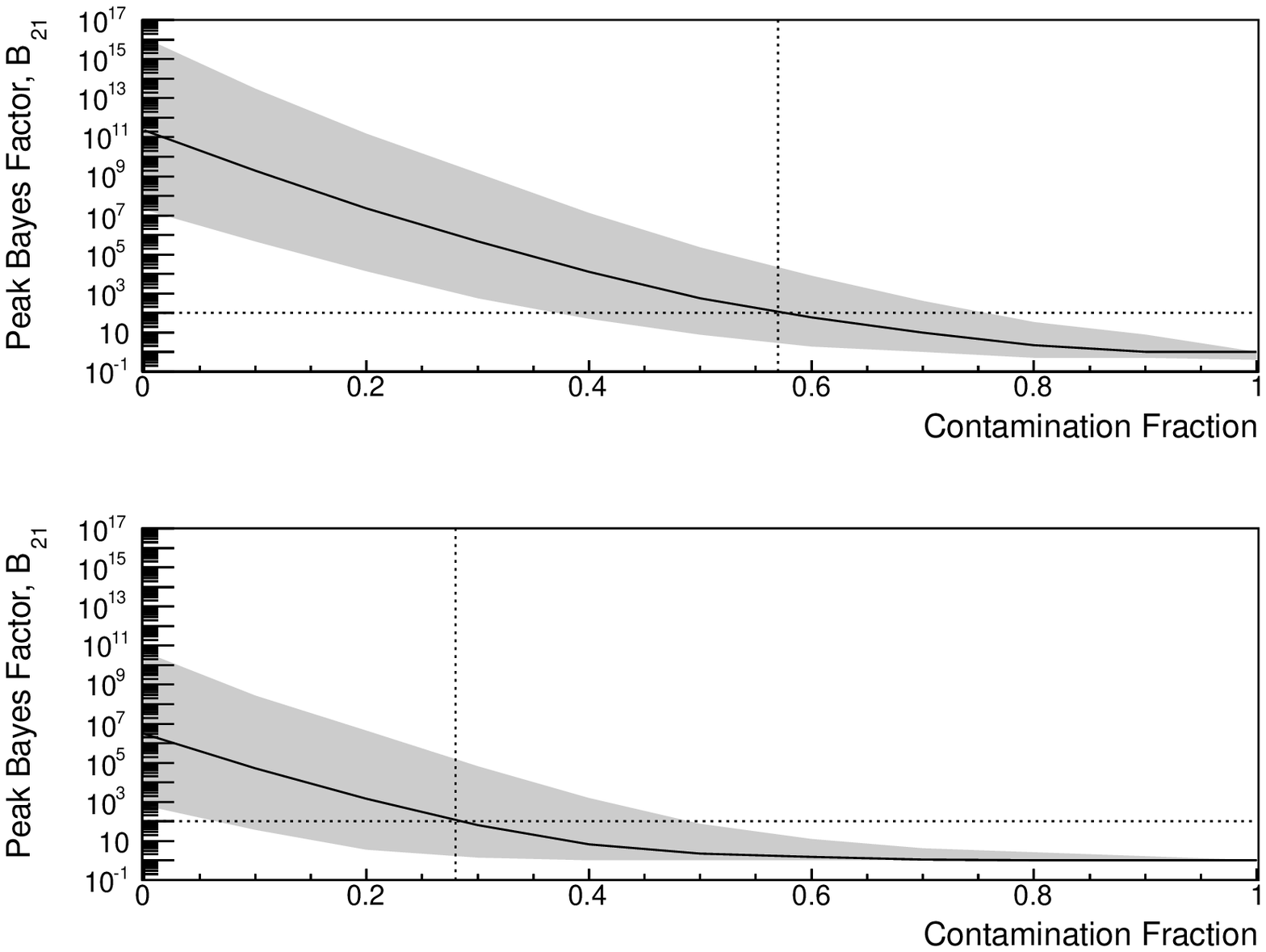}
\caption{Same as Fig.\,\ref{fig:single_broken_5a_reduced}, but for $N_\text{tot}=43557$.
\label{fig:single_broken_5c_reduced}}
\end{figure}

In reality, the region around the source will contain not only source events,
but also background events whose energy distribution follows the all-sky energy
spectrum.  The fraction of background events in the source bin is difficult to
predict as the source flux is not known.  To study the effect, we repeat the
previous analysis assuming that a fraction of the events in the source bin are
background, which we will refer to as the contamination level.  A contamination 
level of 0 means that all events in the source bin are source events, and a
contamination level of 1 means that all events are background events.
Fig.\,\ref{fig:single_broken_5a_reduced} shows the maximum Bayes factor (scanned 
over $E_\mathrm{min}$) as a function of contamination level for $N_\text{tot}=14519$, 
assuming that the source region contains 5\,\% of all events.  Again, we also 
show the results for a data set twice and three times as large
(Fig.\,\ref{fig:single_broken_5b_reduced} and
Fig.\,\ref{fig:single_broken_5c_reduced}, respectively).

The figures indicate that with current Auger statistics, method A can potentially
differentiate spectra at a level of $B_{21}>100$ if the contamination level 
is less than 25\,\%, whereas method B cannot differentiate the spectra for any
level of contamination.  For twice and three times the current data,
method A can differentiate the spectra for contamination levels less than 47\,\% 
and 57\,\%, respectively, and method B for contamination levels of 11\,\% and
28\,\%, respectively.  The difference in the discrimination power of the two methods
is quite substantial, indicating the advantage provided by an accurate knowledge 
of the exposure to the source region.

The Pierre Auger Observatory is scheduled to take data for at least another
decade.  Our studies suggest that in the next few years, as the data increase,
the methods presented here will reach a sensitivity that
enables us to study differences in the spectral shape as a function of sky
position.  A study of the region around a potential source could reveal
significant differences in the energy spectrum compared to the rest of the
sky.

\section{Comparison to a $\chi^{2}$ Test}

In this section, we compare the posterior probability of the one-parent
hypothesis derived from the Bayesian analysis to the tail probability 
obtained from a straightforward two-sample $\chi^2$ test.  We use the 
simple example of the single and broken power law spectrum
from Section\,\ref{subsec:broken}.  The $\chi^2$ test statistic is
\begin{equation}
\chi^2=\frac{1}{n_1 n_2} 
\sum_{i=1}^{m}\frac{(n_2 n_{1i}-n_1 n_{2i})^2}{n_{1i}+n_{2i}}~~,
\label{eq:chisquare}
\end{equation}
where $n_{1i}$ and $n_{2i}$ are the counts for spectrum 1 and 2 in
bin $i$, $n_1$ and $n_2$ are the total number of events in data set 
1 and 2, and $m$ is the number of bins~\citep{Fisher:1922}.  The 
statistic in Eq.\,\ref{eq:chisquare} has approximately a 
$\chi^2_{m-1}$ distribution for samples of sufficient size.  
Whether or not this requirement is met needs to be carefully checked 
for each application, in particular for the comparison of cosmic ray 
spectra at the high-energy tail where the number of events is bound 
to be small.  Assuming that the statistic in Eq.\,\ref{eq:chisquare} 
follows a $\chi^2_{m-1}$ distribution, the $p$-value of the null 
hypothesis that the two spectra are the same above $\log(E/\mathrm{eV})=18.61$ 
is $p=4.4\times 10^{-18}$.

Note that the $\chi^2$ in Eq.\,\ref{eq:chisquare} has scaling constants that
adjust for unequally-sized data samples, so the appropriate Bayesian method for
comparison is method B.  It gives a probability of $9.6\times 10^{-12}$ 
that the compared sets derive from the same parent spectrum. 

\begin{figure}
\epsscale{1.0}
\plotone{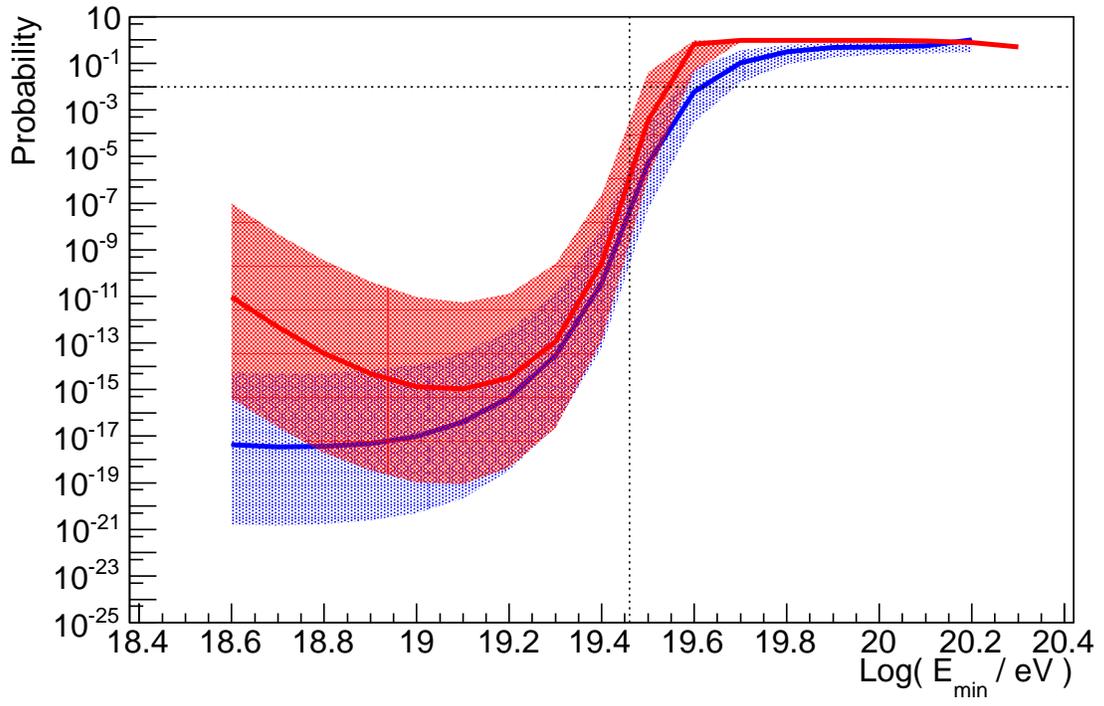}
\caption{Posterior probability of the one-parent hypothesis (red), calculated using
method B, and $\chi^2$ probability (blue) as a function of the lower energy threshold 
$E_\text{min}$ for a comparison between a single power law and broken power law.  
\label{fig:chisquare}}
\end{figure}

The posterior probability and the $\chi^2$ probability are plotted as a 
function of $E_\text{min}$ in Fig.\,\ref{fig:chisquare}.  Both show the same general 
dependence of the probability on $E_\text{min}$.  At face value, the $\chi^2$ 
test gives consistently lower probabilities for all choices of $E_\text{min}$,
which in turn implies that the $\chi^2$ test can resolve smaller differences 
in the energy spectra.  However, our studies show that this is due to the fact
that the test statistic in Eq.\,\ref{eq:chisquare} exhibits considerable 
deviations from the theoretical $\chi^2_{m-1}$ distribution because 
the necessary condition ($n_{1i}\gg1$ and $n_{2i}\gg1$ in each bin) is violated 
in the high-energy tail of the spectrum.
Even for data sets ten times the size of the current Auger 
event sample, the statistics are not sufficient in the high-energy tail of the 
spectrum.  As a result, the use of $\chi^2$ inflates the significance of the 
difference between the sets, and the $\chi^2$ test cannot be applied.  The Bayes 
factor, which assumes Poisson uncertainties in all bins, is not affected 
by this problem and is therefore the appropriate statistical test for comparisons
of energy spectra at the highest energies. 

We also note that at least some of the difference between the $\chi^2$ test
results and the Bayesian method can be attributed to the fact that the 
Bayes factor gives a posterior probability, while the $\chi^2$ gives a 
tail probability.  Tail probabilities are known to be biased 
against the null hypothesis by a factor of at least 10 with respect to 
posteriors~\citep{Sellke:2001}.

\section{Outlook}

The study of the energy spectrum of ultra-high energy cosmic rays at different
parts of the sky is a powerful tool to search for the sources of cosmic rays.
It can supplement direct searches for the sources, which are typically based on 
the statistical analysis of the arrival direction distribution of cosmic rays. 
The direct searches have proven difficult and results are inconclusive 
so far, even with the size and quality of the current generation of cosmic ray 
detectors~\citep{Abreu:2010zzj,Abbasi:2008md}.  However, the cosmic ray data 
set is quickly reaching a size where studies of the shape of the energy spectrum
as a function of sky location can give additional insight into the location and
nature of cosmic ray sources.  The Bayesian method described in this paper 
has several advantages that are important for the comparison of spectra of
ultra-high energy cosmic rays.  It allows for the calculation of the posterior 
odds ratio in favor of the two-parent hypothesis over the one-parent hypothesis,
and it allows for non-Gaussian uncertainties in the data.

An important future application for this analysis is the study of the energy
spectrum in the vicinity of potential sources within the GZK sphere.  With a
data set of about two to three times the size of the last published Auger data
set, the Bayes factor method developed here is sensitive to the difference
between the all-sky cosmic ray energy spectrum and an unattenuated power law
spectrum expected if the source spectrum shows no intrinsic cutoff.

\acknowledgments

This work is supported by the National Science
Foundation under contract number NSF-PHY-0855300.

\bibliography{ms}

\clearpage

\clearpage

\clearpage

\clearpage

\clearpage

\clearpage 

\clearpage

\clearpage

\clearpage

\end{document}